\documentclass[10pt,twocolumn,letterpaper]{article}

\usepackage[pagenumbers]{cvpr} 

\usepackage{graphicx}
\usepackage{amsmath}
\usepackage{amssymb}
\usepackage{booktabs}
\usepackage{adjustbox}
\usepackage{xcolor}

\usepackage[pagebackref,breaklinks,colorlinks]{hyperref}

\usepackage[capitalize]{cleveref}
\crefname{section}{Sec.}{Secs.}
\Crefname{section}{Section}{Sections}
\Crefname{table}{Table}{Tables}
\crefname{table}{Tab.}{Tabs.}

\begin{document}

\title{Handheld Burst Super-Resolution Meets Multi-Exposure Satellite Imagery}

\author{Jamy Lafenetre \qquad Ngoc Long Nguyen \qquad Gabriele Facciolo \qquad Thomas Eboli\\ 
{}\\
Universit\'e Paris-Saclay, ENS Paris-Saclay, CNRS, Centre Borelli, 91190, Gif-sur-Yvette, France\\}

\maketitle

\begin{abstract}
Image resolution is an important criterion for many applications based
on satellite imagery. In this work, we adapt a state-of-the-art
kernel regression technique for smartphone camera burst super-resolution to satellites.
This technique leverages the local structure of the image to
optimally steer the fusion kernels, limiting blur
in the final high-resolution prediction, denoising the image, and recovering
details up to a zoom factor of 2.
We extend this approach to the multi-exposure case
to predict from a sequence of multi-exposure low-resolution
frames a high-resolution and noise-free one.
Experiments on both single and multi-exposure scenarios show
the merits of the approach. Since the fusion is learning-free, the 
proposed method is ensured to not {\em hallucinate} details,
which is crucial for many remote sensing applications.
\end{abstract}

\section{Introduction}

Remote sensing
is an important research field on which are based practical applications
such as natural disaster detection or ecological evaluations.
For each application, image resolution and the 
signal-to-noise ratio (SNR) are two important criterions in practice.
However, the resolution is limited by the aperture
of the telescope, and additional noise further reduce the image
quality, resulting in
low-resolution (LR) and noisy frames.
To circumvent these issues, multi-image
super-resolution (SR) algorithms reconstruct the
underlying high-frequencies spanned in the 
aliasing artifacts~\cite{tsai84multiframe}.
Furthermore, combining multiple images leverages both
spatial and temporal denoising~\cite{dabov07video}.

Nowadays, the best visual accuracy is achieved by
algorithms based on deep learning ~\cite{nguyen21sr, nguyen22hdr}. They significantly
outperform traditional image fusion strategies
based on classical kernel regression~\cite{anger20act}
or inverse problem solvers~\cite{farsiu04fast} both in
speed and visual accuracy.
Notwithstanding, they are not silver-bullet solutions.
First, efficient training of neural networks (NNs) requires very large
amounts of carefully collected supervisory data. 
For image processing
tasks, it translates to perfectly-aligned 
LR burst/HR noise-free target pairs, whose collection
is extremely challenging, especially in remote sensing.
Self-supervised learning (SSL) \cite{nguyen22hdr} have recently 
circumvented this issue, but NNs are known to 
under perform
when they are deployed on tasks not seen during training.
Since synthetic data are limited (hard to model
parallax or occlusions for instance), networks that
perform well on the evaluation data may fell
in practice because of imperceptible details in the images
of the real-world scenario.
Second, NNs are notorious to
{\em hallucinate} details in the HR
prediction. Such details that may sneak from the
training data during evaluation are for 
many applications out-of-question artifacts,
which limit the domains
where NNs can be safely deployed.

In this work, we follow a different trend by adapting to
remote sensing the recent learning-free burst SR approach of~\cite{wronksi19handheld}, proposed for personal
photography.
It  consists  first in aligning the raw frames of a
burst to a reference one with block-matching and 
Lucas-Kanade iterations~\cite{lucas81iterative},
and second in merging the frames into
a HR and denoised image using kernel regression.
The kernels are steered with respect to a
structure tensor~\cite{takeda07kernel}
that retains the details next to the edges and corners
and denoises the flat regions. Structure-adeptness
of the structure tensor
is particularly suited for remote sensing since many objects
such as buildings have regular details such as 
salient edges that must
be restored differently from the flat areas
like fields or the sea. 

However, since the sequences taken by satellites like Planet's SkySat
may have various exposures, with jitter in the exposure
coefficients~\cite{nguyen22hdr}, we cannot expect the approach of \cite{wronksi19handheld} to be a drop-in replacement
of the existing art for remote sensing SR.
First, multi-exposure frame registration is an especially challenging problem~\cite{ward03mtb,lecouat22hdr}, for which the motion
model of \cite{wronksi19handheld} designed for single-exposure is not adapted.
Second, the jitter in the exposure measurements leads
to artifacts in classical kernel regression if
no correction is applied~\cite{nguyen22hdr}.

We address these issues with two fixes: (i) we plug
the NN of \cite{nguyen22hdr} to estimate the optical flow
from variously-exposed frames and show it is accurate
enough when combined with the robustness weight of \cite{wronksi19handheld}, and (ii) we follow
the base-detail decomposition strategy of \cite{nguyen22hdr}
and apply the kernel regression strategy of \cite{wronksi19handheld} on the detail layers of the LR frames,
handling the jitter to the bases. We show that the fusion
technique is flexible-enough to incorporate such decomposition.
Note that with this approach, the final reconstruction is achieved 
by a learning-free module, ensuring that possible 
hallucinations in the predicted optical flow barely 
have consequences in the HR and noise-free prediction. The proposed 
method combines the advantages of learned robust alignment 
for both single and multi-exposure cases, and 
hallucination-free high-quality reconstruction of an 
HR image, all with GPU-accelerated implementations. 
This practical hybrid technique is suitable for a wide 
range of remote sensing applications.

Our contributions are summarized as follows:
\begin{itemize}
    \item We present and adapt the handheld burst SR
    algorithm of \cite{wronksi19handheld} to satellite SR 
    imagery for zoom factors between 1 and 2 and possibly
    important noise levels.
    \item We include the flow estimators and base-detail
    strategy of \cite{nguyen22hdr}, and add a new weight penalizing exposure to adapt the technique to the
    multi-exposure case.
    \item We evaluate the method on both synthetic and real data to illustrate its flexibility and merits. In particular, it copes with the NN-based technique
    from \cite{nguyen22hdr} and exceed the performance
    of the kernel regression of \cite{anger20act}.
\end{itemize}

\section{Related work}

Most approaches for multi-frame SR focus on the single-exposure
case. The idea behind combining multiple frames is to
detect and leverage the aliasing caused by the integration on the sensor that
contains fragments of the original high-frequency content~\cite{tsai84multiframe}.
This has been historically solved 
by accumulating frames in a shift-and-add (SA) strategy~\cite{keren88subpixel, farsiu04fast},
by solving inverse problems~\cite{farsiu06multiframe, baker02limits}, or via kernel regression~\cite{anger20act, wronksi19handheld, takeda07kernel, lafenetre23handheld, briand18low}.
These approaches reconstruct a signal with a higher pixel
density, and thus containing details beyond the Nyquist rate
of the sensor. However they are blurry because of 
the blurring inherent to interpolators, and the reconstruction
of the lens point-spread function (PSF)~\cite{baker02limits}. As a result, a subsequent
deblurring~\cite{farsiu06multiframe, anger20act} or sharpening~\cite{eboli22polyblur} is thus required in practice to produce the
final image.

Despite being successful in many applied fields, including
remote sensor, the state of the art is nowadays dominated
by deep learning, \eg~\cite{bhat21reparam}. These approaches are notorious for
the large amounts of high-quality supervisory data they require
for training, yet such high-quality LR/HR image pairs are hardly
obtainable in the context of satellites. To circumvent this
issue, Nguyen \etal\cite{nguyen21sr} train a CNN
with a self-supervised loss. However, and despite improved
visual accuracy over the handcrafted counterparts, these methods
may hallucinate details, which is not compatible with
many remote sensing applications. In contrast, we propose
to adapt the kernel regression technique from \cite{wronksi19handheld},
which is: efficiently parallelized on a GPU, 
signal-adaptive, robust to motion and noise, and learning-free, thus
ensuring no hallucination while providing high-quality results.

Multi-exposure imagery is another important family of multi-image
methods that are highly relevant for remote sensing. In high-dynamic
range (HDR) imagery, taking sequences of images at different
exposures with limited dynamic range, and fusing them together
results in a new HDR one~\cite{debevec97hdr}. Neural networks
may be trained on bursts of bracketed LR frames to jointly address 
HDR and SR reconstruction~\cite{lecouat22hdr}. 
Nguyen \etal\cite{nguyen22hdr} propose such an approach for remote 
sensing, again trained in a self-supervised manner. In particular,
they train a CNN to predict the optical flow between two frames
differently exposed, a very challenging problem in the HDR literature~\cite{ward03mtb, zimmer11freehand}.
In this work, we adapt the kernel regression approach of \cite{wronksi19handheld} to satellite bracketed bursts by plugging
the optical flow CNN of \cite{nguyen22hdr} to align
the bracketed frame, and include typical HDR weights~\cite{tsin01ccd, granados10optimalhdr} to these kernels to reconstruct high-quality
HDR and SR satellite images.

\begin{figure*}
    \centering
    \includegraphics[trim=0 475 0 0,clip,width=0.85\textwidth]
{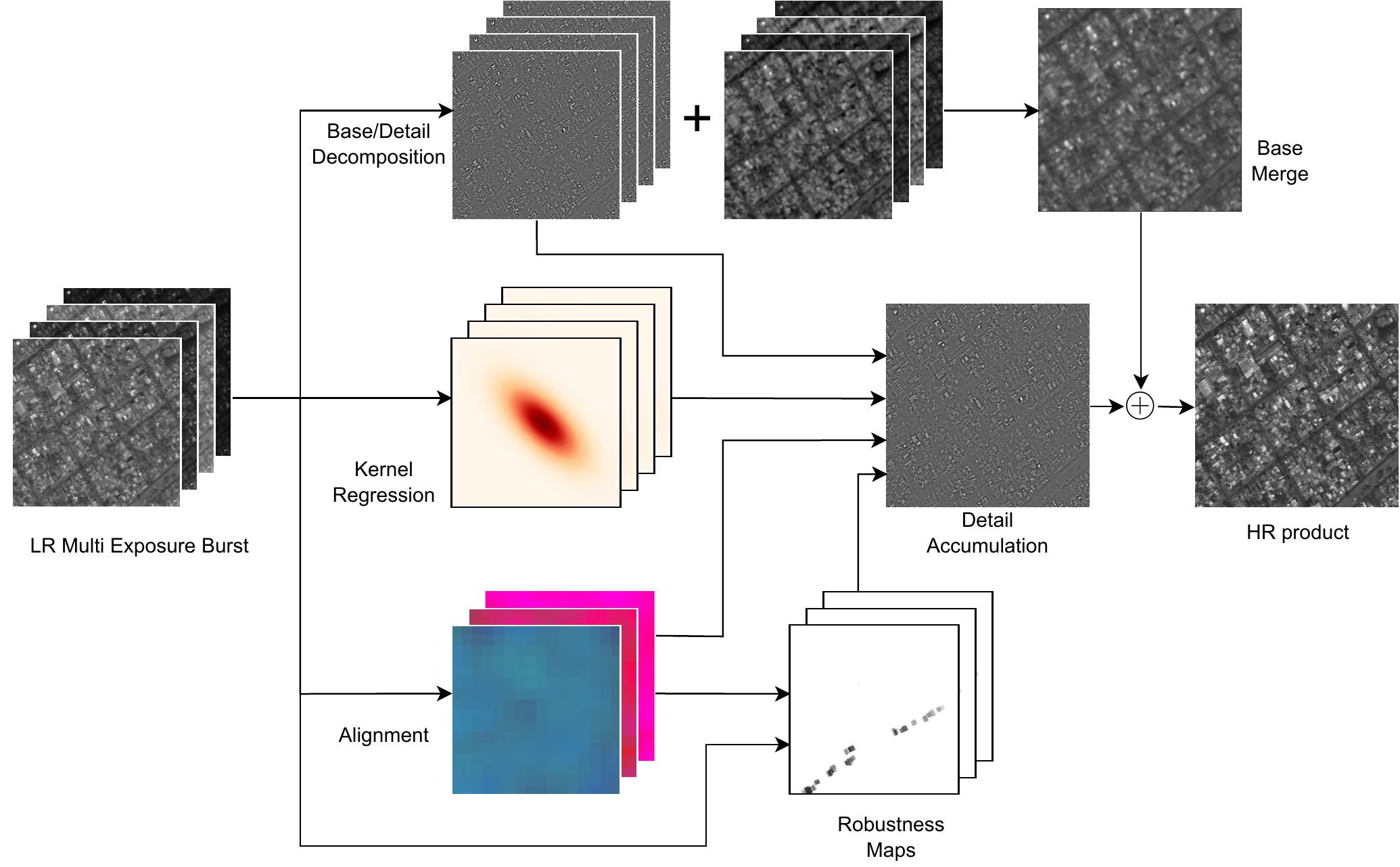}
    \caption{General logic of our pipeline. Merge kernels are estimated, images are aligned and robustness maps are determined. 
    A base/detail decomposition is performed on the burst, and the details of the bursts are accumulated using normalized convolution. The bases are merged, upscaled to HR, then fused with the HR details.}
    \label{fig:approach}
\end{figure*}

\section{Approach}

\subsection{Multi-Exposure Kernel Regression}
The data of the problem are the $n$ LR raw frames $J_n$ and
corresponding exposures $t_n$ ($n=1,\dots,N$). Our
goal is to predict a single HR frame $I$ aligned
with a reference in the LR sequence. In the single-exposure case, any frame is equally valid, but in the multi-exposure
case, each frame may have a different SNR depending on exposure~\cite{hasinoff10noiseoptimal}, and saturated areas
impossible to align. We consider the frames for which the pixel saturation rate is below a hand-fixed threshold (such as $ 5\%$), 
and chose the most exposed among them. If no frame is below the threshold, then the least exposed frame is chosen.

The proposed approach is shown in Figure~\ref{fig:approach}.
It illustrates the several stages that we describe in
this section.
First, we normalize the raw LR images by the exposure coefficients, for all $n$:
\begin{equation}\label{eq:exposurenorm}
    H_n = J_n / t_n,
\end{equation}
and compute the corresponding saturation masks
via a threshold arbitrary set to 0.99 as in~\cite{lecouat22hdr}:
\begin{equation}\label{eq:saturation}
    M_n = J_n < 0.99.
\end{equation}
From these normalized images, we first compute the
optical flows between the frames $H_2,\dots,H_n$ and $H_1$
that will be used during the fusion. In theory any method
may work, but in Sec.~\ref{sec:experiments} we 
show that the FNet model of \cite{nguyen22hdr} is the most
performing and reliable one for remote sensing, significantly
more than
the patchwise inverse compositional algorithm (ICA) iterations~\cite{baker01ica}
used in \cite{lafenetre23handheld} and more suited to the personal photography.
Let us name FNet $f$. The $n$ $(n=1,\dots,N)$ flow images are obtained as:
\begin{equation}\label{eq:flow}
    F_n = f(H_1, H_n),
\end{equation}
with the convention that $F_1$ is 0 (no motion). The
``Alignment'' module in Fig.~\eqref{fig:approach}
thus returns $N$ flow images $F_1,\dots,F_N$.

Following Nguyen \etal\cite{nguyen22hdr}, we proceed
by decomposing the $n$ LR frames $H_n$ into base and detail layers
to be robust to exposure jitter:
\begin{subequations}\label{eq:basedetaildecomp}
    \begin{align}
        B_n &= H_n * G_\sigma,\\
        D_n &= H_n - B_n,
    \end{align}
\end{subequations}
with $G_\sigma$ a Gaussian filter of variance $\sigma^2$,
set to 1 in our experiments.
The HR estimated base
is simply the accumulation of the LR base images
upsampled by bilinear interpolation, and warped
on the reference as:
\begin{equation}\label{eq:base}
    B = b\left(\frac{1}{N}\sum_{n=1}^N W(H_n, F_n)\right),
\end{equation}
where $W$ is the warp function that aligns with 
bilinear interpolation the frame $H_n$ with respect to the flow $F_n$, and $b$ is the bilinear upscaling operation 
by a factor $s$.
This is valid because the base images only contain the 
low-frequencies up to the cut-off of the Gaussian filter $G_\sigma$. Equation~\eqref{eq:base}
corresponds to the ``Base Merge'' module
in Fig.~\ref{fig:approach}.

More care is given to the recovering of the HR details that contains
frequencies beyond the Nyquist rate. It is achieved
with the steerable kernel regression strategy of \cite{wronksi19handheld}, and reads:
\begin{equation}\label{eq:detail}
    D(x,y) = \frac{\sum_n \sum_{(p,q)\in\mathcal{N}} k_n(p,q) D_n(p,q)}{\sum_n \sum_{(p,q)\in\mathcal{N}} k_n(p,q)},
\end{equation}
where $\mathcal{N}$ is the $3\times3$ neighborhood
of pixels in each LR frame that are the closest to the 
location $(x,y)$ on the HR grid (details in \cite{lafenetre23handheld}).
Equation~\eqref{eq:detail} is accounted by the module ``Detail Accumulation'' in Fig.~\ref{fig:approach}.
The $k_n$'s are computed as the multiplication of three weights:
\begin{equation}
    k_n(p,q) = w_n(p,q)r_n(p,q)h_n(p,q),
\end{equation}
where $w_n$ is a geometric weight, $r_n$ is a robustness weight that rejects mobile objects and artifacts, and $h_n$ is an HDR weight that gives more importance to frames with better exposure.
The  first two weights come from~\cite{wronksi19handheld}
and are those corresponding to burst SR. The latter, dubbed $h_n$,
is a contribution of this work to handle the multi-exposed
frames.
The final image is the summation of the HR predicted
base and detail layers:
\begin{equation}\label{eq:finalimage}
    I = B + D.
\end{equation}
In what follows we detail all the intermediate
results to obtain the predicted details HR layers.

\subsection{Description of the weights}
An overview of these weights
is presented in \cite{wronksi19handheld}, and implementation details
are disclosed in \cite{lafenetre23handheld}.

\paragraph{Geometric weight}
The geometric weight barely differs from the original paper. It corresponds to the ``kernel regression''
module in Fig.~\ref{fig:approach}. It consists
in steerable kernels adapted to the local geometry: edges, corners
or flat areas.
Briefly it reads
\begin{equation}
    w_n(p,q) = \exp\left(-\frac{1}{2} 
    \begin{bmatrix}
        x-p \\
        y-q
    \end{bmatrix}^\top 
    \Omega^{-1} 
    \begin{bmatrix}
        x-p \\
        y-q
    \end{bmatrix}\right),
\end{equation}
where $\Omega$ is the locally adaptive covariance matrix. 
$\Omega$ is shaped by the hyperparameters $k_{detail}$ and $ k_{denoise}$, and $D\in[0, 1]$, measuring the amount of details (presence of noise and/or textures). 
It is estimated using the
local structure tensor~\cite{takeda07kernel, wronksi19handheld}. 
We build $\Omega$ in the eigen basis $P$ of the local structure tensor 
as follows:
\begin{equation}
    \Omega = 
    P\begin{bmatrix}
    k_1^2 & 0\\
    0 & k_2^2
    \end{bmatrix}P^\top.
\end{equation}
In the case where both direction have
equal energy in the structure tensor like
on a corner or a flat region,
\ie isotropic behavior, $k_1$ is equal to
$k_2$, and reads:
\begin{equation}\label{eq:kernels_iso}
    k_1 = (1-D)k_{detail} + Dk_{denoise}.
\end{equation}
Conversely, when on an edge, where a direction
has much more energy than the other one,
$k_1$ and $k_2$ are different and stretch
the kernel along the edge:
\begin{subequations}\label{eq:kernels_aniso}
    \begin{align}
        k_1 &= (1-D)[0.5k_{detail}] + Dk_{denoise},\\
        k_2 &= (1-D)[4k_{detail}] + Dk_{denoise}.
    \end{align}
\end{subequations}
In this anisotropic case, $k_{detail}$ is made 
smaller for the normal direction to the edge
to avoid collecting pixels across it, thus reducing
blur, and is enlarged along the edge direction
to increase spatial denoising without blurring.
This is a unique feature of the steerable
kernels, not implemented in~\cite{anger20act}.

\begin{figure}
    \centering
    \includegraphics[trim=0 600 340 0,clip,width=0.75\linewidth]
{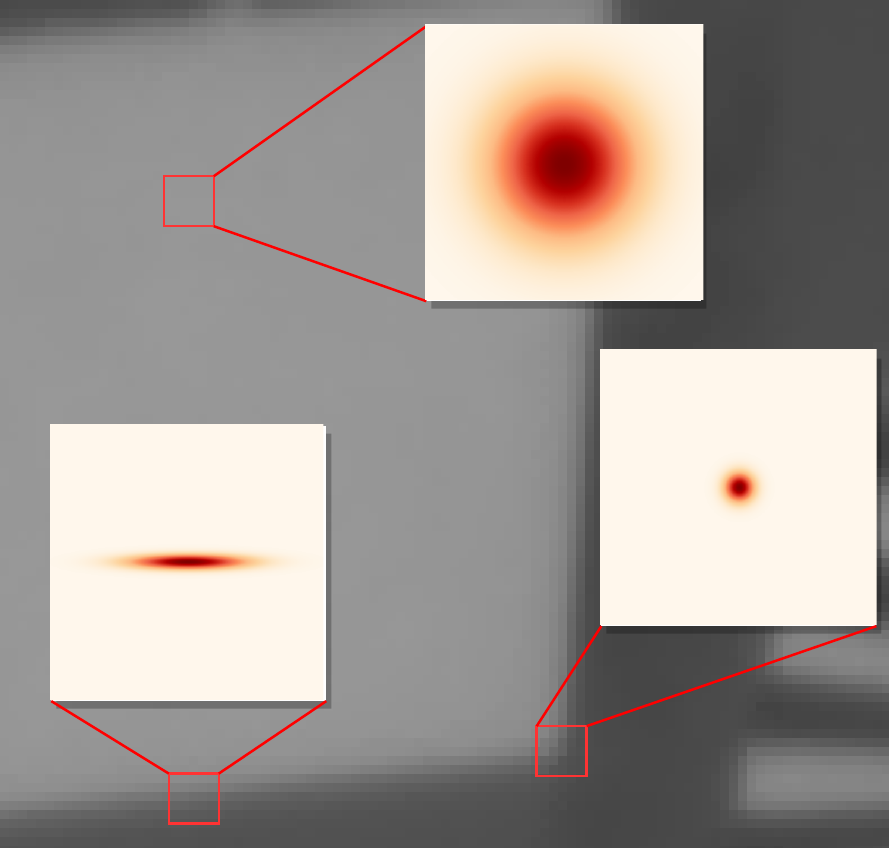}
    \caption{Illustration of the adaptive kernels of our method. A large isotropic kernel is used for areas without details, 
    and a narrow isotropic kernel is used for areas such as corners. A stretched kernel is used for edges.}
    \label{fig:kernels_illustration}
\end{figure}

In both cases, following the value of $D$, the kernel $w_n$
is made larger to denoise, or smaller to
prevent blurring of the corners and edges.
The amount of spatial denoising and detail
conservation is controlled by $k_{denoise}$
and $k_{detail}$ that are two hyper-parameters
automatically set by the estimated SNR score.
More details on $k_1$ and $k_2$ can be found in ~\cite{wronksi19handheld, lafenetre23handheld}.
Overall, this approach is a data-adaptive
way to combine images, a merit of CNNs but in a learning-free manner.
We show examples of 
steered kernels for a flat area, an edge and a corner
in a real image in Figure~\ref{fig:kernels_illustration}.

\paragraph{Robustness weight}

Robustness is tailored to make plausible natural images from
everyday's life scenes with deformable objects and numerous occlusions.
This is far more challenging than the satellite imagery case where
the common assumption is to assume static scenes. 
In this paragraph, all the quantities are pixelwise but we drop this dependency for the sake of conciseness.
The robustness is based around the ratio $d/\sigma$ at a LR pixel location $(p,q)$, 
where $d$ measures the difference between a pixel and its matching position on the reference frame, and $\sigma$ represents the local variance~\cite{wronksi19handheld}.
Let $p_1$ and $p_n$ be two corresponding patches in the images $H_1$
and $W(H_n, F_n)$. A preliminary step consists in computing the mean
color difference $d_n = \| \mu(p_1) - \mu(p_n) \|_2$ with $\mu$ the mean
function, and the standard deviation of $p_1$, dubbed $\sigma_1$.
Since these statistics are computed from few pixels on $3\times3$ 
neighborhoods, they might be too noisy to be used directly. Instead,
they are corrected with the expected values $d_s$ and $\sigma_s$
obtained via simulations as:
\begin{equation}
    d = \frac{d^2}{d^2 + d_s^2}d\quad\text{and}\quad\sigma =\max(\sigma_s,\sigma_1).
\end{equation}
Wronski \etal simulate the expected values $\sigma_s, d_s$ by performing Monte-Carlo on the clipped Poisson-Gaussian noise model, 
for every ISO and binned brightness levels, since $p_1$ and $p_n$ have the same brightness in their framework. 
After this correction, higher values of the ratio 
indicate that an area may be prone to artifacts, 
whereas low ratios characterize safe to merge pixels: the aliased
details necessary for SR, and the noisy flat areas important for
efficient temporal denoising.
In the single-exposure setting, the robustness coefficient is then obtained using the ratio of the corrected values:
\begin{equation}\label{eq:localrobustness}
    r_n(p,q)= \text{clip}\left(s \exp\left(-\frac{d^2}{\sigma^2}\right) - t, 0, 1\right).
\end{equation}
The scaling factor $s$ is a function of the magnitude of the local
optical flow. If the flow is too large, the risk of misalignments
artifacts is more important. Therefore the patch
is deemed unreliable, and $s$ is set to 2, otherwise to the much larger value of 12.
The threshold $t$ is set to 0.12 as in \cite{lafenetre23handheld}.
In our case, it also prevents to aggregate patches for which FNet
may have returned a false prediction, \eg hallucination.
Lastly, the robustness weights in \eqref{eq:localrobustness} are
 pooled on a $5\times 5$ local neighborhood to share the worst-case
confidence. This pessimistic strategy further prevents the 
accumulation of possible artifact-prone patches. 

However, in the multi-exposure case
the normalized images $H_1$ and $H_n$, and thus the patches $p_1$
and $p_n$ may have roughly the same brightness but their SNRs
remains different. This penalizes the ratio $d/\sigma$ even if
the two patches are visually similar, thus unnecessarily discarding
important frames for denoising and SR.
Therefore, we would need to simulate $d_s$ and $\sigma_s$ for every exposure ratio, every ISO and every binned brightness. However, we do not need to adapt to many camera settings as in \cite{wronksi19handheld}, and can therefore simulates the curves for the single ISO gain of the satellite, thus making this approach tractable.
If we had access to the exact noise characteristics of the SkySat 
satellite, we could also include in the simulation of $d_s$
and $\sigma_s$ the dependency on the exposure of the 
noise profile~\cite{hasinoff10noiseoptimal,granados10optimalhdr}.
This would further improve the quality of the robustness. 
An example of robustness map detecting mobile
objects in a real sequence is show in Fig.~\ref{fig:carmotion}, 
in the experiments section.

\paragraph{HDR weight}
The additional HDR weight compared to \cite{wronksi19handheld},
gives more importance to the well-exposed frames, and filters out the
saturated areas and the darker regions with the lowest SNRs.
This weight differs with the robustness weight in many ways. First, it relies only on a single frame, whereas the robustness is defined for image pairs via $d$. Second, the HDR weight may discard the reference if better
exposed frames are available, especially for saturated areas in the reference.
We use the weight of \cite{tsin01ccd}, since a reliable 
estimate of the noise standard deviation $\sigma_s$ has already been computed for 
the robustness weight.
This weight reads
\begin{equation}\label{eq:hdrweights}
  h_n(p,q) =  \frac{t_n}{\sigma_s(p, q)}M_n(p,q),
\end{equation}
where $\sigma_s(p, s)$ is the noise standard deviation estimated for the mean brightness of the pixel located in $(p, q)$. 
Ideally, it should be obtained using the noise curve specifically estimated for the exposure $t_n$, but we use the same curve for all 
frames since only a single noise profile was available to us. Note that we could have also used the local
statistics of the robustness stage to compute HDR weights based
on local estimates of the SNR~\cite{granados10optimalhdr, hasinoff10noiseoptimal}, a common
practice in the HDR community.
We have noted in our experiments that those of \eqref{eq:hdrweights}
were enough to achieve satisfactory results.

\section{Experiments}
\label{sec:experiments}

We base our implementation on the official 
codes from \cite{lafenetre23handheld} and \cite{nguyen22hdr}.
Our code will be made publicly available.

\begin{table}[]
    \centering
    \begin{tabular}{lcccc} 
        \toprule
           & SE & ME-0\% & ME-5\% & ME-20\% \\
        \midrule
         ICA-P & 52.24 & 52.46 & 52.29 & 50.87 \\
         ICA-G & \textbf{53.78} & \textbf{53.58} & \underline{53.34} & \underline{51.76} \\
         FNet & \underline{53.57} & \underline{53.41} & \textbf{53.40} & \textbf{53.34} \\
        \bottomrule
    \end{tabular}
    \caption{Comparison of different registration techniques plugged to the steerable kernel regression module. We report the average PSNR for 
    200 synthetic bursts of 15 $256\times256$ LR frames in the single-exposure (SE) and multi-exposure (ME) settings. For the latter, we follow \cite{nguyen22hdr} and inject jitter
    in $\{0,5,20\}$\% to the exposures to measure robustness of flow
    estimation to such practical artifacts. For ME, we use BD decomposition for the merge, in order to restrain the effect of jitter to the flow estimation.}
    \label{tab:alignment}
\end{table}

\subsection{Alignment}
We first evaluate the quality of the output image using three different registration
methods. We compare the patch-wise ICA algorithm shipped with the code
of \cite{lafenetre23handheld} and adapted for personal photography,
the global ICA algorithm used in \cite{anger20act}, adapted to satellite
imagery since most motions across images are that of the satellite itself,
and the CNN dubbed FNet from \cite{nguyen22hdr} trained in an
end-to-end manner via self-supervised learning on multi-exposure
synthetic bursts.

We report in Table~\ref{tab:alignment} the average PSNR
estimated on 200 simulated bursts of size $15\times256\times256$ for both single-exposure (SE) and multi-exposure (ME) with 3 jitter rates as in \cite{nguyen22hdr}.
The patchwise ICA (dubbed ICA-P) can suffer from instabilities and does not perform as good as ICA global (ICA-G) and FNet. For low jitter rates, global ICA performs better since the synthetic dataset was generated using a global translation model. Yet, FNet falls shortly behind, and performs consistently as the jitter rate rises, contrary to both ICA method for which the PSNR drops.  Note a drop of about 2dB for ICA-G compared to FNet
for the most severe jitter on the exposure ratios. This suggets
that FNet is more robust for general ME scenarios.
However in the case of SE, the three methods are in the same
ballpark but we note that FNet and ICA-P better handle mobile
objects than ICA-G in practice. Because the kernel regression
methods are modular-enough to be plugged with most registration
techniques, we show that the choice of the alignment is problem-dependent.

\begin{figure}[t]
    \centering
    \adjustbox{max width=\linewidth}{
    \begin{tabular}{ccc}
        \begin{subfigure}[t]{0.32\linewidth}
            \centering
            \includegraphics[trim=25 10 15 10,clip,width=\textwidth]{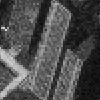}
            \caption{Reference.}
        \end{subfigure} &
        \begin{subfigure}[t]{0.32\linewidth}
            \centering
            \includegraphics[trim=25 10 15 10,clip,width=\textwidth]{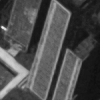}
            \caption{Low.}
        \end{subfigure} &
        \begin{subfigure}[t]{0.32\linewidth}
            \centering
            \includegraphics[trim=25 10 15 10,clip,width=\textwidth]{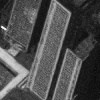}
            \caption{High.}
        \end{subfigure} 
    \end{tabular} 
    }
    \caption{Illustration of the difference between narrow and wide kernels for noisy bursts (standard deviation of $3/255$). Wide kernels allow for a better denoising while narrow kernels allow for a better recovery of high frequencies.}
    \label{fig:kerneltradeoff}
\end{figure}

\subsection{Impact of noise}\label{noisesection}
Raw measurements are degraded by noise coming from both the nature
of light and the electronics~\cite{hasinoff10noiseoptimal}. A 
super-resolution method should thus be robust to several signal-to-noise (SNR)
ratios to deliver high-quality results. In \cite{wronksi19handheld},
the magnitude of the steerable kernels is further shaped by the SNR in the 
raw images via two parameters $k_{detail}$ and $k_{denoise}$: 
the lower the SNR the larger the kernels to achieve
better spatial denoising.

We show in Table~\ref{tab:ablationnoise} the performance of the
SR approach for single-exposure for 4 sets of 200 synthetic bursts
of size $15\times256\times256$, each one degraded with white Gaussian
noise of standard devation in  $\{0.5, 1, 2, 3\}/255$. We test
three sets of $(k_{detail},k_{denoise})$ to monitor their efficiency
on satellite images: Low $(0.33px, 1.65px)$ (the default parameters in \cite{lafenetre23handheld}), Medium $(0.24px, 0.96px)$ and High $(0.15px, 0.45px)$.
Each set of parameter is particularly adapted to certain SNRs. Medium
is an all-purpose setting to handle both small and important noise instances
whereas the default parameter Low is only adapted to the least
favorable case, as expected.
We illustrate in Figure~\ref{fig:kerneltradeoff} the impact
of choosing among the ``Low'' and `High'' sets of parameters.
The one specialized for low SNRs tends to overblur the image, whereas
that for higher SNRs may reconstruct correlated noise when achieving
SR. Notwithstanding, note that the method may retrieve very fine
details with noisy measurements such as the stripes on the building
with the ``High'' setting.

\begin{figure*}[ht]
    \centering
    \adjustbox{max width=\textwidth}{
    \begin{tabular}{ccccc}
    \begin{subfigure}[t]{0.2\linewidth}
        \centering
        \includegraphics[trim=50 25 55 25,clip,width=\textwidth]{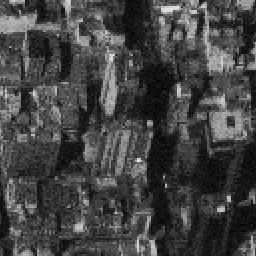}
        \caption{Reference.}
    \end{subfigure} &
    \begin{subfigure}[t]{0.2\linewidth}
        \centering
        \includegraphics[trim=50 25 55 25,clip,width=\textwidth]{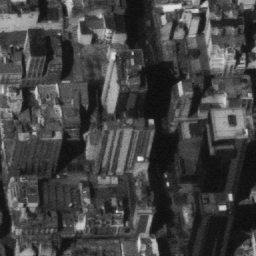}
        \caption{SA.}
    \end{subfigure} &
    \begin{subfigure}[t]{0.2\linewidth}
        \centering
        \includegraphics[trim=50 25 55 25,clip,width=\textwidth]{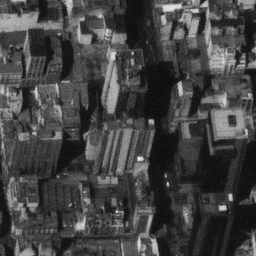}
        \caption{ACT~\cite{anger20act}.}
    \end{subfigure} &
    \begin{subfigure}[t]{0.2\linewidth}
        \centering
        \includegraphics[trim=50 25 55 25,clip,width=\textwidth]{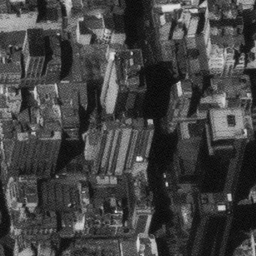}
        \caption{DSP~\cite{nguyen22hdr}.}
    \end{subfigure} &
    \begin{subfigure}[t]{0.2\linewidth}
        \centering
        \includegraphics[trim=50 25 55 25,clip,width=\textwidth]{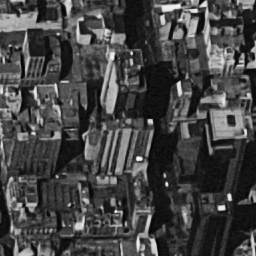}
        \caption{Ours.}
    \end{subfigure}
    \end{tabular} 
    }
    \caption{Example of joint denoising and SR for $N=15$ frames and noise of standard deviation of $3/255$. Our approach automatically steers 
    the kernels to produce a HR noise-free image. In contrast, DSP~\cite{nguyen22hdr} returns a HR image with correlated noise. Better seen on a computer screen.}
    \label{fig:noisequalitative}
\end{figure*}

We show in Table~\ref{tab:singleexposurenoise} a comparison with an implementation of
shift-and-add (SA), and two state-of-the-art approaches: ACT-spline (dubbed ACTS)~\cite{anger20act}
and  DSP, the CNN from \cite{nguyen22hdr}. We set the Gaussian noise
level to 3/255, leading too poor SNR in the measurements. 
Our approach with the ``Low''
setting better handles such instances of noise. 
Note that the CNN was not retrained, and may therefore underperform.
Figure~\ref{fig:noisequalitative} shows a qualitative comparison
of the same panel of methods for a sequence with low SNR. ACTS~\cite{anger20act}
is not designed to jointly address denoising and SR, and thus correlates
the noise. DSP~\cite{nguyen22hdr} also correlate the noise but
restores sharp details. Our approach in the other hand may efficiently remove
important noise while recovering a lot of details.

\begin{table}[]
    \centering
    \begin{tabular}{lccccc}
        \toprule
        Noise std. & 0.5 & 1 & 2 & 3 &\\
        \midrule
        Low & 51.69 & 51.66 & \underline{50.57} & \textbf{48.73} &\\
        Medium & \underline{53.08} & \underline{52.82} & \textbf{50.68} & \underline{47.98} & \\
        High & \textbf{55.07} & \textbf{53.78} & 49.08 & 45.35 & \\
        \bottomrule
    \end{tabular}
    \caption{Comparison of the mean PSNR estimated for 200 bursts of 15 $256\times256$ LR frames in the single-exposure (SE) setting, for different noise std. Three sets of 
    kernel parameters $(k_{detail}, k_{denoise})$ are considered : Low $(0.33px, 1.65px)$, Medium $(0.24px, 0.96px)$ and High $(0.15px, 0.45px)$. 
    Each parameter set outperforms the 2 others for at least one noise level.}
    \label{tab:ablationnoise}
\end{table}

\begin{table}[t]
    \centering
    \begin{tabular}{lcc}
        \toprule
         & PSNR & SSIM \\
        \midrule
        SA~\cite{nguyen21sr} & \underline{46.81} & \underline{0.995}\\
        ACTS~\cite{anger20act} & 45.46 & 0.993\\
        DSP~\cite{nguyen22hdr} & 42.52 & 0.985\\
        Ours & \textbf{48.73} & \textbf{0.996}\\
        \bottomrule
    \end{tabular}
    \caption{Single-exposure SR $\times 2$ with stack size $N=15$. Average PSNR on 200 bursts with Gaussian noise of standard deviation of $3/255$.}
    \label{tab:singleexposurenoise}
\end{table}

\subsection{Handling mobile objects}

\begin{figure}[t]
    \centering
    \adjustbox{max width=\linewidth}{
    \begin{tabular}{cc}
        \begin{subfigure}[t]{0.48\linewidth}
            \centering
            \includegraphics[trim=10 10 10 10,clip,width=\textwidth]{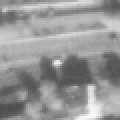}
            \caption{Reference.}
        \end{subfigure} &
        \begin{subfigure}[t]{0.48\linewidth}
            \centering
            \includegraphics[trim=10 10 10 10,clip,width=\textwidth]{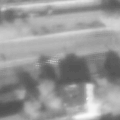}
            \caption{DSP~\cite{nguyen22hdr}.}
        \end{subfigure} \\
        \begin{subfigure}[t]{0.48\linewidth}
            \centering
            \includegraphics[trim=10 10 10 10,clip,width=\textwidth]{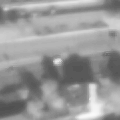}
            \caption{Ours.}
        \end{subfigure} &
        \begin{subfigure}[t]{0.48\linewidth}
            \centering
            \includegraphics[trim=10 10 10 10,clip,width=\textwidth]{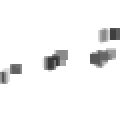}
            \caption{Robustness mask.}
        \end{subfigure} \\
    \end{tabular} 
    }
    \caption{Illustration of the robustness mask of our model. The white dot represents a vehicle partially occluded by trees, moving forward on a road. 
    The darkest points on the accumulated robustness mask indicate areas where frames are massively rejected due to scene motion, and where the accumulation mostly relies on the reference frame.}
    \label{fig:carmotion}
\end{figure}

The scenes in synthetic bursts are by construction static; it is therefore not relevant to evaluate performance with the robustness mask on such data, especially considering that the flow estimation is very reliable. However, we show in Figure~\ref{fig:carmotion} a crop from a real sequence
featuring moving cars on a road. The prediction of DSP~\cite{nguyen22hdr}
splatters the car along the road , whereas our result retains the car.
This is crucial for several applications of remote sensing such as 
surveillance. This difference is explained by the fact that
our method includes a robustness mask that attributes a map of confidence 
to each frame: when every frame but the reference is rejected, 
our method is a mere upsampling technique. Yet, it retains the moving details
in contrast to the CNN. We also show in Figure~\ref{fig:carmotion}
the accumulation of the confidence score that we call ``robustness mask''.
This image gives hints on the mobile objects and misalignments, and can
be used in practice to understand the behavior of our approach, whereas
the black-box CNN cannot be diagnosed in case of failure.

\begin{table}[t]
    \centering
    \begin{tabular}{lccc}
        \toprule
         & $N=5$ & $N=10$ & $N=15$ \\
        \midrule
        SA & 49.14 & 51.83 & 53.11 \\
        ACTS~\cite{anger20act} & 48.88 & 51.64 & 52.93 \\
        DSP~\cite{nguyen22hdr} & \textbf{51.21} & \underline{52.61} & \underline{53.49} \\
        Ours & \underline{50.79} & \textbf{52.74} & \textbf{53.78} \\
        \bottomrule
    \end{tabular}
    \caption{Single-exposure SR $\times 2$ with varying stack size $N$. Average PSNR on 200 bursts of size $N$ varying in $\{5,10,15\}$, and noise of standard deviation of $1/255$.}
    \label{tab:singleexposure}
\end{table}

\subsection{Single-exposure validation}

We quantitatively evaluate our approach with the panel
composed of SA, ACTS~\cite{anger20act}, and DSP~\cite{nguyen22hdr}.
We generate 200 bursts of size $15\times256\times256$ from HR 
crops of the L1A satellite dataset with the protocol
detailed in \cite{nguyen21sr}: we blur the image with
a Gaussian filter with standard deviation of $0.3$, translate the other
frames than the reference with a subpixel shift in the Fourier domain,
decimate by 2 with nearest interpolation these images, and 
ultimately add Gaussian noise of standard deviation of $1/255$.
Since the noise level of this test set is low, we select the
``High'' set of values for $(k_{detail}, k_{denoise})$ as previously
discussed.

Table~\ref{tab:singleexposure} shows the average PSNR scores 
for the panel of methods we consider, and so for three
burst size: $N$=5, 10, and 15. It can be seen that for $N=5$
images, we fall short by less than 1dB to DSP~\cite{nguyen22hdr}. We rank second for this burst size, and above the other kernel regression technique ACTS~\cite{anger20act} by a margin of 1.6dB. 
However for larger burst sizes, we rank first with margins of 0.13dB and 0.29dB over DSP. We are thus in the same ballpark as deep learning
for these more practical values of $N$, validating our approach.
We also keep important margins of about 1dB over ACTS. Our adaptive
kernels better preserve the details such as the edges and corners
whereas ACTS is equivalent to kernel regression with isotropic
kernels~\cite{briand18low}, thus blurring details.

We have also noted during our experiments that our method is mostly as good as the deepnet to handle instances of parallax next to skyscrapers in real-world images. This shows the merits of our method to urban scenes.

\begin{table}[]
    \centering
    \begin{tabular}{lccccc}
        \toprule
        & Time~(ms/burst) & Peak mem.~(GB) \\
        \midrule
        SA & $49.5 \pm 2.7$ & 3.1 \\
        DSP~\cite{nguyen22hdr} & $548.3\pm 22.8$ & 10.8\\
        Ours (ICA) & $129.7\pm 14.4$ & 2.0\\
        Ours (FNet) & $118.6\pm 10.1$ & 2.4\\
        \bottomrule
    \end{tabular}
    \caption{Execution time per burst (s/burst) of size $15\times256\times256$ pixels on a single NVIDIA RTX 3090 graphic card. We benchmark our method for the patchwise 
    ICA alignment, since an efficient GPU implementation had already been designed for \cite{lafenetre23handheld}, as well as with FNet flows.
    }
    \label{tab:speed}
\end{table}

Lastly, we report in Table~\ref{tab:speed} the
average running time for 200 bursts of size $15\times256\times256$. Our approach, while 
relying on a non-official reimplementation of 
the handheld method~\cite{lafenetre23handheld}, 
is faster than the 
CNN from \cite{nguyen22hdr}, and much more memory-efficient, showcasing its
practicality. This showcases that the bulk of the computations in \cite{nguyen22hdr} are attributed to the fusion stage, that is as accurate with
our learning-free, but for a much smaller computational cost.
Since Wronski \etal\cite{wronksi19handheld} claim to process 
in 100ms on a 2018's smartphones 
a dozen of 12 Megapixel raw photographs with their own non-released implementation, speed improvements are expected with a better engineered
implementation than the non-official one of \cite{lafenetre23handheld}.
Note that the original method merges the frames sequentially in order to fit a low memory device; the runtime could therefore be improved further at the cost of a heavier memory usage by merging frames simultaneously.
We do not report the running time
for ACTS~\cite{anger20act} since a CPU-only code exists; 
It is slower
by several order of magnitudes compared to the other methods.
We remark that SA has a larger peak memory usage despite being much simpler than 
the steerable kernel strategy. This is because the code from \cite{nguyen22hdr} 
process all the images of the stack in parallel, whereas we sequentially process the
frames, thus keeping the memory usage low.

\subsection{Multi-exposure validation}

We compare the performance of the proposed approach
for SR of multi-exposed sequences.
Table~\ref{tab:multiexposure} shows the performance of different algorithms, evaluated on the synthetic dataset. 
We evaluated in our panel the kernel regression bases techniques, \ie \cite{anger20act} and ours, with and without
the base-detail (BD) decomposition proposed in \cite{nguyen22hdr}. DSP is run with the BD decomposition.
We observe that the BD decomposition is mostly beneficial for high jitter rates, but ensures a consistent PSNR over a wide range of jitter values. 
Our method ranks second behind DSP, with the advantage of a smaller memory footprint and computational cost. It is consistently better than ACTS,
confirming that data-adaptivity leads to
better accuracy.
Remark that this evaluation setting is the most favorable to DSP, since the network was trained for this exact noise profile. It was shown in 
Section~\ref{noisesection} that our method remains competitive over a broad range of noise intensities, and can therefore outperform NNs on images
with different SNRs.

\begin{table}[t]
    \centering
    \begin{tabular}{lccc}
        \toprule
         & Exp.~0\% & Exp.~5\% & Exp.~20\% \\
        \midrule
        ACTS~\cite{anger20act} & 53.19 & 51.35 & 44.78 \\
        ACTS (B.D)~\cite{nguyen22hdr} & 52.79 & 52.71 & 50.97 \\
        SA & 53.42 & 53.00 & 49.60 \\
        DSP~\cite{nguyen22hdr} & \textbf{55.54} & \textbf{55.54} & \textbf{55.49} \\
        Ours & \underline{54.53} & 52.98 & 46.07\\
        Ours (B.D) & 53.41 & \underline{53.40} & \underline{53.34}\\
        \bottomrule
    \end{tabular}
    \caption{Multi-exposure SR $\times2$. Average PSNR on
    200 bursts of $N=15$ frames and exposure ratio jitter in $\{0,5,20\}\%$.}
    \label{tab:multiexposure}
\end{table}

\subsection{Limitations}

We have observed during evaluation on certain
real images that some details are not as well restored as
ACTS~\cite{anger20act} and DSP~\cite{nguyen22hdr}.
When switching to structure tensor-independent narrow isotropic kernels, these pixel-thin details
are restored. We posit it comes from the computation
of the gradients since \cite{wronksi19handheld} average the
gradients of several neighboring pixels to be robust to noise. We have observed in practice that for these small details, the dominant eigenvalue of the tensor was not as high as expected, thus favoring too-large kernels
where details should be retained instead.
This is acceptable in \cite{wronksi19handheld} because
of the image resolution, whereas the details in our LR frames may be tiny and lead to
incorrect gradients, resulting in overblurring of certain details.
Better tuning of the hyper-parameters or task-specific image gradients should alleviate this issue. 

\section{Conclusion}

We embedded the fast and efficient steerable kernel regression approach from~\cite{wronksi19handheld} for single-exposure burst SR
into a hybrid scheme for multi-exposure SR in 
the remote sensing context.
We combined this image-fusion module with two modules from \cite{nguyen22hdr}: the
neural network for optical flow,
trained on real images to handle the challenging problem of varying exposures,
and the base-detail decomposition strategy to handle jitter in the exposure 
coefficient.
This combination is a fast and interpretable blend of learnable
flow and handcrafted image fusion, taking the best of the two worlds
for estimating robust complex motions, and merging the frames with the
guarantee to never {\em hallucinate} details.
The latter is a key criterion for many practical remote sensing applications,
and we thus believe that the proposed approach is perfectly suited for both
academia and the industry.
Experiments for both single-exposure and multi-exposure frames empirically validate our
approach.

\paragraph{Acknowledgements:} This work was partly financed by the DGA Astrid Maturation project SURECAVI ANR-21-ASM3-0002, the Office of Naval 
research grant N00014-17-1-2552, and the ANR project IMPROVED ANR-22-CE39-0006-04. This work was performed using HPC resources from GENCI–IDRIS (grant 2022-AD011012453R1). 
Centre Borelli is also a member of Université Paris Cité, SSA and INSERM. We thank J\'er\'emy Anger for the fruitful discussions.

{\small
\bibliographystyle{ieee_fullname}
\bibliography{egbib}
}

\end{document}